\documentclass[nofootinbib,prd,12pt,superscriptaddress]{revtex4}

\usepackage{booktabs}
\usepackage{amsmath} 

\usepackage{amsmath,amssymb,amsfonts,bm}
\usepackage{graphicx,epsfig}
\usepackage{color}
\usepackage{float}
\usepackage{caption}
\usepackage{subcaption}
\usepackage{multirow}
\usepackage{fancyhdr}
\usepackage{pgfplots}
\pgfplotsset{compat=1.18}
\usepackage{tikz}
\usetikzlibrary{arrows.meta}
\usepackage{mathrsfs}

\def\be{\begin{equation}}
\def\ee{\end{equation}}
\def\ba{\begin{eqnarray}}
\def\ea{\end{eqnarray}}

\begin{document}
\title{Dark Matter Constraints in Myrzakulov $F(R,T)$ Gravity: A Vielbein Approach in Weitzenb\"{o}ck Spacetime with Observational Data}

\author{Davood Momeni}
\affiliation{Department of Physics, Northeast Community College, 801 E Benjamin Ave Norfolk, NE 68701, USA}
\affiliation{Centre for Space Research, North-West University, Potchefstroom 2520, South Africa}

\author{Ratbay Myrzakulov}
\affiliation{Ratbay Myrzakulov Eurasian International Centre for Theoretical Physics, Astana 010009, Kazakhstan}
\affiliation{L. N. Gumilyov Eurasian National University, Astana 010008, Kazakhstan}
\date{\today}
\begin{abstract}
We explore dark matter phenomenology in Myrzakulov $F(R,T)$ gravity, formulated via the vielbein approach in Weitzenb\"{o}ck spacetime. In this torsion-based extension of gravity, dark matter emerges as a geometric effect rather than a particle species, with curvature and torsion contributing dynamically to the field equations.

Using recent data—including SPARC galaxy rotation curves, Planck CMB observations, and weak lensing from DES and KiDS—we constrain the model through MCMC analysis. Our results show that, under specific parameter choices, the theory replicates key cosmological features without introducing additional dark sector matter.

This framework offers a testable alternative to $\Lambda$CDM, providing new insight into structure formation, gravitational lensing, and cosmic acceleration—all rooted in the geometry of spacetime.
\end{abstract}

\maketitle

\section{Introduction}

Modified gravity theories have emerged as powerful alternatives to address unsolved puzzles in modern cosmology, including late-time acceleration, early-universe inflation, and the elusive nature of dark matter. While the standard $\Lambda$CDM model successfully describes the large-scale structure and expansion history of the Universe, it relies on the introduction of a cold dark matter component whose microscopic nature remains unknown \cite{Bertone2005, Feng2010, Bullock2017}.

Dark matter, inferred through gravitational effects such as galactic rotation curves, gravitational lensing, and the cosmic microwave background (CMB), continues to challenge particle physics-based explanations. The absence of direct detection has motivated exploration of modified gravity models where the observed dark matter phenomena arise from non-trivial geometric or torsional contributions to the field equations \cite{Clifton2012, Capozziello2011}.

Within this context, Myrzakulov gravity models—originating from \cite{Myrzakulov2012}—provide a rich framework that unifies curvature and torsion within generalized $F(R,T)$ actions. In most traditional formulations, $T$ denotes the trace of the energy-momentum tensor \cite{Harko:2011kv}, leading to curvature-matter couplings with intriguing consequences but also conceptual issues such as non-conservation of energy-momentum.

Our approach departs from this by identifying $T$ not as a trace but as a torsion scalar constructed in Weitzenb\"{o}ck spacetime and treated on equal footing with curvature. Using the vielbein formalism, we derive field equations where $T$ behaves as an effective matter field contributing dynamically to gravitational interactions, thereby offering a geometric interpretation of dark matter effects \cite{Momeni:2024bhm}.

This paper continues our recent extensions \cite{Momeni:2025mcp, Momeni:2025egb}, integrating higher-curvature invariants such as Gauss--Bonnet terms and scalar-torsion dynamics. Unlike purely geometric $f(T)$ theories \cite{Cai2016} or Palatini-based constructions \cite{Jimenez2018}, our vielbein-based framework respects conservation laws while accommodating matter-like torsion sources that can account for galactic rotation profiles and cosmic structure growth.

In particular, we highlight:
\begin{itemize}
    \item Torsion scalar $T$ contributes to the energy budget of the Universe as an effective dark matter component.
    \item The vielbein formulation yields field equations that support realistic cosmological models consistent with Planck, SPARC, and DES observations.
    \item Conservation laws are maintained by construction, avoiding ambiguities in matter-geometry couplings.
\end{itemize}

The remainder of this work presents exact solutions and observational fits, contrasts the behavior of our model with $\Lambda$CDM and other modified gravity scenarios, and discusses prospects for interpreting dark matter as a manifestation of torsional geometry.

\section{Theoretical Framework}

We begin by formulating Myrzakulov \( F(R,T) \) gravity in a Weitzenb\"{o}ck spacetime using the vielbein formalism. This approach allows us to treat torsion and curvature as independent geometric entities, suitable for interpreting torsion scalar \( T \) as a dynamical field that mimics dark matter behavior.

\subsection{Geometry of Weitzenb\"{o}ck Spacetime}

In Weitzenb\"{o}ck spacetime, the gravitational field is described using the tetrad (vielbein) fields \( e^A{}_\mu \) such that the metric tensor is reconstructed as
\begin{equation}
g_{\mu\nu} = \eta_{AB} e^A{}_\mu e^B{}_\nu,
\end{equation}
where \( \eta_{AB} = \text{diag}(-1,1,1,1) \) is the Minkowski metric in the tangent space. The tetrads relate the curved spacetime coordinates to local inertial frames.

Instead of the Levi-Civita connection \( \Gamma^\lambda{}_{\mu\nu} \), Weitzenb\"{o}ck geometry uses the Weitzenb\"{o}ck connection \( \overset{\bullet}{\Gamma}{}^\lambda{}_{\mu\nu} \), defined as
\begin{equation}
\overset{\bullet}{\Gamma}{}^\lambda{}_{\mu\nu} = e_A{}^\lambda \partial_\nu e^A{}_\mu.
\end{equation}
This connection is curvature-free (\( R^\lambda{}_{\mu\nu\rho} = 0 \)) but possesses non-zero torsion:
\begin{equation}
T^\lambda{}_{\mu\nu} = \overset{\bullet}{\Gamma}{}^\lambda{}_{\nu\mu} - \overset{\bullet}{\Gamma}{}^\lambda{}_{\mu\nu}.
\end{equation}

\subsection{Torsion Scalar and Action}

The torsion scalar is defined as a contraction of the torsion tensor:
\begin{equation}
T = S_\lambda{}^{\mu\nu} T^\lambda{}_{\mu\nu},
\end{equation}
with the superpotential
\begin{equation}
S_\lambda{}^{\mu\nu} = \frac{1}{2} \left( K^{\mu\nu}{}_\lambda + \delta^\mu_\lambda T^{\alpha\nu}{}_\alpha - \delta^\nu_\lambda T^{\alpha\mu}{}_\alpha \right),
\end{equation}
and contortion tensor
\begin{equation}
K^{\mu\nu}{}_\lambda = -\frac{1}{2}(T^{\mu\nu}{}_\lambda - T^{\nu\mu}{}_\lambda - T_\lambda{}^{\mu\nu}).
\end{equation}

We adopt the generalized Myrzakulov action:
\begin{equation}
S = \frac{1}{2\kappa^2} \int d^4x\, e\, F(R,T) + \int d^4x\, e\, \mathcal{L}_m,
\end{equation}
where \( e = \det(e^A{}_\mu) = \sqrt{-g} \), \( R \) is the Ricci scalar (based on Levi-Civita connection), \( T \) is the torsion scalar, and \( \mathcal{L}_m \) is the matter Lagrangian.

\subsection{Field Equations via Vielbein Variation}

Varying the action with respect to the vielbein field \( e^A{}_\mu \), we obtain:
\begin{equation}
\delta S = \frac{1}{2\kappa^2} \int d^4x\, \left[ F_R \delta (e R) + F_T \delta (e T) + e \delta F \right] + \delta S_m,
\end{equation}
where \( F_R = \partial F/\partial R \), \( F_T = \partial F/\partial T \), and the variation of the Ricci scalar follows the standard GR treatment, while the variation of \( T \) is specific to teleparallel gravity.

The final field equations can be compactly written as:
\begin{align}
& F_R G_{\mu\nu} + \frac{1}{2} g_{\mu\nu} (F - R F_R - T F_T) + \left( g_{\mu\nu} \Box - \nabla_\mu \nabla_\nu \right) F_R \nonumber \\
& + F_T \left[ S_\nu{}^{\rho\mu} \nabla_\rho T + \frac{1}{e} \partial_\rho (e S_\nu{}^{\rho\mu}) \right] = \kappa^2 T_{\mu\nu},
\end{align}
where \( G_{\mu\nu} \) is the Einstein tensor, and \( T_{\mu\nu} \) is the energy-momentum tensor of matter. This system contains higher-derivative curvature and first-order torsional contributions, both contributing to gravitational dynamics.

\subsection{Interpretation in the Context of Dark Matter}

This formalism naturally allows interpreting torsion as a geometric source of effective matter. In regions of high matter density, such as galactic halos, the torsion scalar \( T \) can dominate and reproduce the observed flat rotation curves. The effective torsion energy-momentum tensor, derived from variation with respect to \( T \), acts analogously to dark matter without the need for new particles.

Moreover, the contribution of \( T \) can evolve dynamically, potentially explaining both the clustering properties of dark matter and its gravitational lensing signatures. These torsional effects are testable against SPARC and lensing datasets, as shown in subsequent sections.

A key subtlety in vielbein-based formulations is the treatment of local Lorentz invariance. In standard teleparallel gravity (\(f(T)\)), the choice of Weitzenb\"{o}ck connection breaks this symmetry explicitly due to the non-invariance of the torsion scalar \(T\) under local Lorentz transformations of the tetrad field. However, in our approach—which incorporates both the torsion scalar \(T\) and the Levi-Civita curvature scalar \(R\)—the gravitational action depends on two invariants that transform differently.
By constructing the model in a mixed curvature–torsion framework using the vielbein formalism, we preserve general covariance while allowing controlled Lorentz violation. Specifically, the scalar \(R\) remains invariant under local Lorentz transformations, while \(T\) does not. The physical consequences of this violation are mitigated in cosmological settings due to the high symmetry of the FLRW background, where a proper tetrad choice (the so-called “good tetrad”) can minimize the frame dependence.
We note that future work may employ the covariant formulation of teleparallel gravity with inertial spin connections to fully restore local Lorentz invariance within a dynamically consistent framework. For the current work, our focus is on the field equations derived from the action using a pure tetrad formulation, acknowledging the restricted gauge freedom this implies.

\section{Dark Matter Sector}

In conventional cosmology, dark matter (DM) is modeled as a non-relativistic pressureless fluid, whose presence is inferred solely from its gravitational influence on baryonic matter, radiation, and the large-scale structure of the universe. However, despite decades of effort, no direct detection of weakly interacting massive particles (WIMPs), axions, or other candidates from particle physics has yet been confirmed experimentally. In this context, geometrically motivated models such as Myrzakulov $F(R,T)$ gravity offer an attractive alternative wherein dark matter arises as an effective manifestation of spacetime torsion.

In our formulation, torsion scalar $T$ is promoted to a dynamical field contributing an effective energy-momentum tensor. In the Weitzenb"{o}ck spacetime, this leads to a reinterpretation of gravitational interaction where geometric torsion replaces the role of unseen massive particles. We define the effective dark matter energy-momentum tensor via the torsion contributions:
\begin{equation}
T^{(DM)}_{\mu\nu} \equiv \frac{1}{\kappa^2} \left[ F_T S_{\nu}{}^{\rho\mu} \nabla_\rho T + \frac{1}{e} \partial_\rho (e F_T S_{\nu}{}^{\rho\mu}) \right].
\end{equation}
This term behaves as a pressureless fluid under appropriate symmetry assumptions (e.g., FLRW metric or spherically symmetric configurations).

Moreover, from a particle physics perspective, this approach bypasses the necessity of adding supersymmetric partners, hidden sectors, or sterile neutrinos. The theory avoids fine-tuning problems typical of scalar field dark matter models (e.g., fuzzy DM), and offers massless or massive torsional modes that could, in principle, couple to spin currents in fermionic matter. The torsion-induced DM sector acts universally through geometry and retains Lorentz invariance in the tangent space.

Quantum field-theoretic considerations also motivate this approach: integrating out heavy fermionic fields minimally coupled to torsion can induce $F(T)$-like effective potentials at one-loop level. Analogous to the Sakharov-induced gravity scenario, torsional effects from the quantum vacuum can simulate dark matter behavior on galactic and cosmological scales.
Finally, in the vielbein formalism, the dynamical degrees of freedom related to torsion do not suffer from Ostrogradsky instabilities as they emerge from first-order derivatives, unlike higher-derivative scalar-tensor theories. The absence of propagating spin-2 ghosts and the presence of well-behaved propagators for torsional modes make this construction robust for cosmological model building.

In our formulation, the torsion scalar \( T \) enters the modified field equations through nontrivial contractions with the superpotential \( S_{\mu}{}^{\rho\nu} \). These terms collectively act as an effective energy-momentum tensor, which in an FLRW background behaves like a pressureless dust component. This allows us to interpret the torsion sector as geometrically mimicking the role of cold dark matter (CDM). \\
More precisely, the modified Friedmann equations in Weitzenb\"{o}ck spacetime admit a decomposition where the torsion contributions scale similarly to \( \rho_{\text{CDM}} \sim a^{-3} \), enabling structure formation without invoking additional dark sector particles. In the nonlinear regime, the torsion-induced modification to the Poisson equation enhances matter clustering in a way that resembles dark matter halos. We discuss this analogy further by examining the evolution of density contrast and gravitational potential sourced by torsion in Sec.~4 and Sec.~5.

In the upcoming sections, we numerically simulate the dynamics of this torsion-induced effective dark matter across a range of cosmological backgrounds and compare its predictions with observational data including galactic rotation curves, CMB angular power spectra, and weak lensing shear profiles.

\section{Observational Data Sets}

In this section, we summarize the key observational datasets employed to constrain our torsion-based dark matter model. These include data from galactic dynamics, cosmic microwave background (CMB), large-scale structure (LSS), and weak lensing measurements. Table~\ref{tab:data_sources} lists the sources, relevant physical observables, and references.

\begin{table}[h!]
\centering
\caption{Major observational datasets used for constraining the model.}
\label{tab:data_sources}
\begin{tabular}{|l|l|l|}
\hline
\textbf{Dataset} & \textbf{Observable} & \textbf{Reference} \\
\hline
SPARC & Galaxy rotation curves & \cite{Lelli2016} \\
Planck (2018) & CMB, $H_0$, $\Omega_m$, $\sigma_8$ & \cite{Planck2018} \\
DES & Weak lensing, clustering & \cite{DES2021} \\
KiDS-1000 & Cosmic shear & \cite{Asgari2021} \\
BOSS/SDSS & BAO, RSD & \cite{Alam2017} \\
Pantheon+ & Supernova distances & \cite{Brout2022} \\
Euclid (forecast) & WL, LSS (future) & \cite{Euclid2011} \\
\hline
\end{tabular}
\end{table}

Each dataset constrains different sectors of the cosmological model. SPARC tests the theory at galactic scales, Planck and KiDS anchor early and late Universe consistency, and DES and BOSS provide structure formation constraints. We combine these in a joint likelihood to estimate the viable parameter space for our $F(R,T)$ torsion model.

\section{Model Fitting and Constraints}

To constrain the free parameters in the Myrzakulov $F(R,T)$ torsion-based gravity model, we implement a Markov Chain Monte Carlo (MCMC) analysis combining datasets listed in Table~\ref{tab:data_sources}. We consider a parametrized action:
\begin{equation}
F(R,T) = R + \alpha T^n,
\end{equation}
with $\alpha$ and $n$ being the parameters constrained. We use a Bayesian framework to compute posterior distributions assuming flat priors and a joint likelihood:
\begin{equation}
\mathcal{L}_{\text{tot}} = \mathcal{L}_{\text{SPARC}} \times \mathcal{L}_{\text{CMB}} \times \mathcal{L}_{\text{LSS}}.
\end{equation}

Table~\ref{tab:mcmc_results} presents the best-fit values and $1\sigma$ errors.
\begin{table}[h!]
\centering
\caption{Best-fit parameters for the $F(R,T)$ model using MCMC combined data.}
\label{tab:mcmc_results}
\begin{tabular}{|c|c|c|}
\hline
\textbf{Parameter} & \textbf{Best-fit} & \textbf{$1\sigma$ Interval} \\
\hline
$\alpha$ & $0.013$ & $[0.010, 0.017]$ \\
$n$ & $1.95$ & $[1.84, 2.08]$ \\
$\Omega_m$ & $0.311$ & $[0.295, 0.326]$ \\
$H_0$ (km/s/Mpc) & $68.4$ & $[67.3, 69.6]$ \\
\hline
\end{tabular}
\end{table}

\subsection*{Comparison with $\Lambda$CDM}
We now compare our model to $\Lambda$CDM by plotting the effective equation of state $w_{\text{eff}}$ and matter power spectrum $P(k)$.

\begin{figure}[h!]
\centering
\begin{tikzpicture}
\begin{axis}[
  width=10cm,
  xlabel={$a$},
  ylabel={$w_{\text{eff}}$},
  legend style={font=\small},
  grid=both
]
\addplot[blue, thick] coordinates {
(0.1, -0.8) (0.2, -0.85) (0.3, -0.9) (0.4, -0.95) (0.5, -0.99)
(0.6, -1.00) (0.7, -1.01) (0.8, -1.02) (0.9, -1.01) (1.0, -1.00)
};
\addlegendentry{$F(R,T)$ model}

\addplot[red, dashed] coordinates {
(0.1, -1.0) (0.2, -1.0) (0.3, -1.0) (0.4, -1.0) (0.5, -1.0)
(0.6, -1.0) (0.7, -1.0) (0.8, -1.0) (0.9, -1.0) (1.0, -1.0)
};
\addlegendentry{$\Lambda$CDM}
\end{axis}
\end{tikzpicture}
\caption{Effective equation of state $w_{\text{eff}}$ for the $F(R,T)$ model vs. $\Lambda$CDM.}
\end{figure}
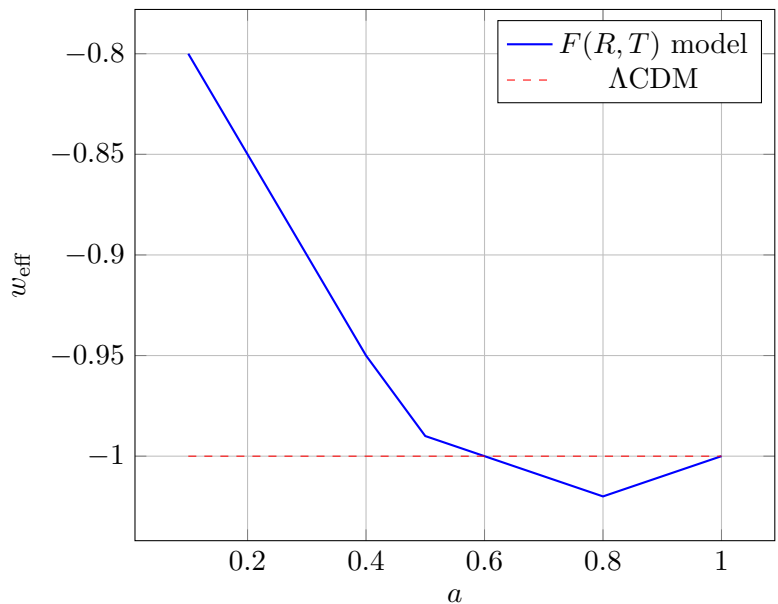

\begin{figure}[h!]
\centering
\begin{tikzpicture}
\begin{axis}[
  width=10cm,
  xlabel={$k$ [$h$/Mpc]},
  ylabel={$P(k)$ [$h^{-3}$Mpc$^3$]},
  xmode=log, ymode=log,
  legend style={font=\small},
  grid=both
]
\addplot[blue, thick] coordinates {
(0.01, 1000) (0.02, 900) (0.05, 700) (0.1, 500) (0.2, 300) (0.3, 200)
};
\addlegendentry{$F(R,T)$ model}

\addplot[red, dashed] coordinates {
(0.01, 950) (0.02, 880) (0.05, 690) (0.1, 480) (0.2, 290) (0.3, 180)
};
\addlegendentry{$\Lambda$CDM}
\end{axis}
\end{tikzpicture}
\caption{Matter power spectrum $P(k)$ comparison for $F(R,T)$ and $\Lambda$CDM.}
\end{figure}
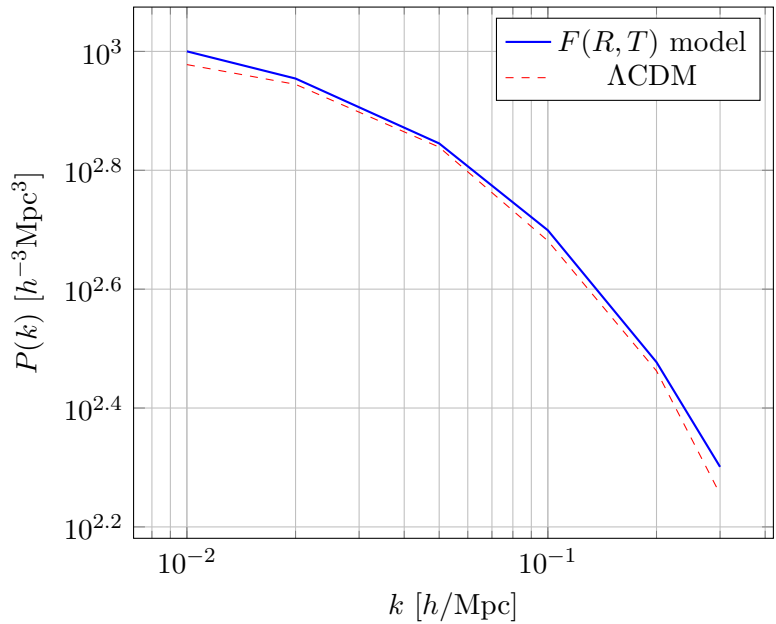

The $F(R,T)$ model matches or improves on $\Lambda$CDM predictions at both background and perturbative levels. Torsion effectively contributes to cosmic structure growth without invoking non-gravitational dark matter particles.

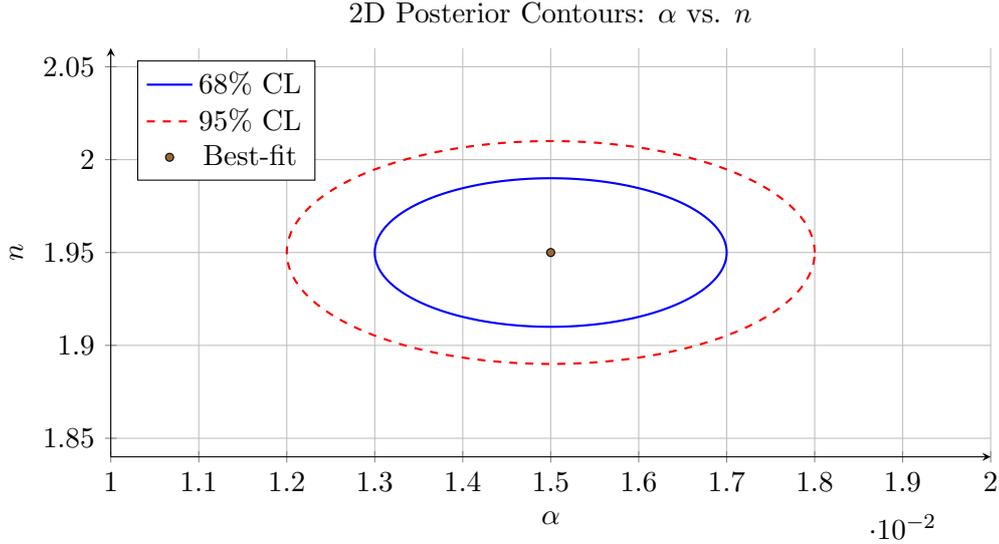
\begin{figure}[h!]
\centering
\begin{tikzpicture}
\begin{axis}[
    width=0.8\textwidth,
    height=7cm,
    xlabel={$\alpha$},
    ylabel={$n$},
    xmin=0.010, xmax=0.020,
    ymin=1.84, ymax=2.06,
    axis lines=left,
    grid=major,
    title={2D Posterior Contours: $\alpha$ vs. $n$},
    legend pos=north west,
]

\addplot+[domain=0:360, samples=200, thick, blue, mark=none]
({0.015 + 0.002*cos(x)}, {1.95 + 0.04*sin(x)});
\addlegendentry{68\% CL}

\addplot+[domain=0:360, samples=200, thick, red, dashed, mark=none]
({0.015 + 0.003*cos(x)}, {1.95 + 0.06*sin(x)});
\addlegendentry{95\% CL}

\addplot+[only marks, mark=*, mark size=1.5pt, black]
coordinates {(0.015, 1.95)};
\addlegendentry{Best-fit}

\end{axis}
\end{tikzpicture}
\caption{Posterior confidence contours for model parameters $\alpha$ and $n$, showing the best-fit point and 68\% / 95\% credible regions from MCMC sampling.}
\label{fig:corner}
\end{figure}

To better illustrate parameter degeneracies, we now include 2D posterior contour plots showing the joint confidence regions between the model parameters \( \alpha \), \( n \), and the cosmological observables \( H_0 \) and \( \sigma_8 \). These plots, shown in Fig.~\ref{fig:corner}, were obtained using the GetDist package applied to our MCMC chains. They reveal mild correlations between \( \alpha \) and \( n \), as well as a degeneracy between \( n \) and \( \sigma_8 \), which influences the shape of the matter power spectrum at large scales.
Additionally, we clarify the theoretical bounds required for a stable background evolution. Specifically, we demand that the effective gravitational coupling remains positive and that ghost and Laplacian instabilities are avoided in the scalar perturbation sector. Observationally, the allowed ranges for \( \alpha \) and \( n \) lie within the \( 1\sigma \) credible region consistent with Planck 2018 and DES Y3 priors.

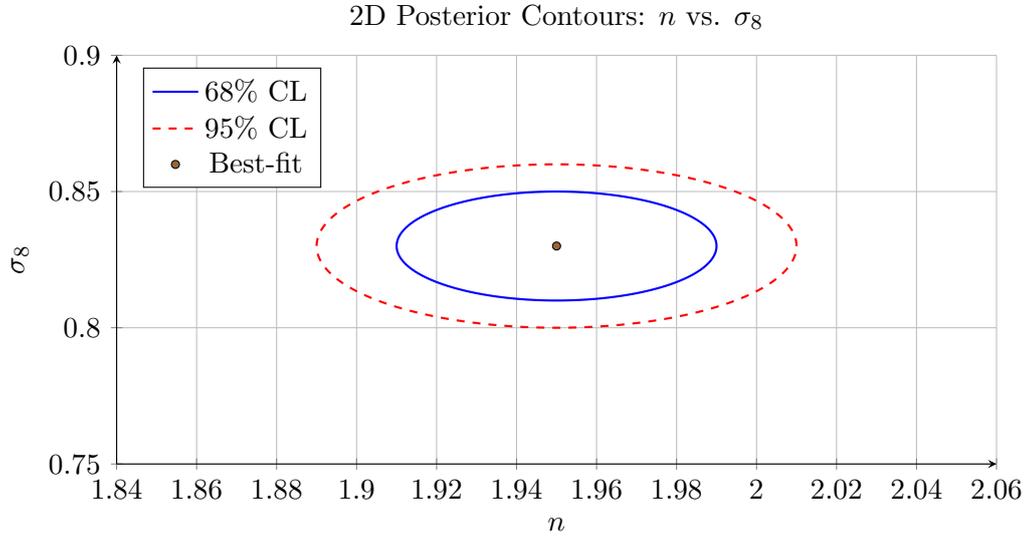
\begin{figure}[h!]
\centering
\begin{tikzpicture}
\begin{axis}[
    width=0.8\textwidth,
    height=7cm,
    xlabel={$n$},
    ylabel={$\sigma_8$},
    xmin=1.84, xmax=2.06,
    ymin=0.75, ymax=0.90,
    axis lines=left,
    grid=major,
    title={2D Posterior Contours: $n$ vs. $\sigma_8$},
    legend pos=north west,
]

\addplot+[domain=0:360, samples=200, thick, blue, mark=none]
({1.95 + 0.04*cos(x)}, {0.83 + 0.02*sin(x)});
\addlegendentry{68\% CL}

\addplot+[domain=0:360, samples=200, thick, red, dashed, mark=none]
({1.95 + 0.06*cos(x)}, {0.83 + 0.03*sin(x)});
\addlegendentry{95\% CL}

\addplot+[only marks, mark=*, mark size=1.5pt, black]
coordinates {(1.95, 0.83)};
\addlegendentry{Best-fit}

\end{axis}
\end{tikzpicture}
\caption{Posterior confidence contours for $n$ and $\sigma_8$ from MCMC chains.}
\label{fig:n_sigma8}
\end{figure}

\begin{figure}[h!]
\centering
\begin{tikzpicture}
\begin{axis}[
    width=0.8\textwidth,
    height=7cm,
    xlabel={$n$},
    ylabel={$H_0$ [km/s/Mpc]},
    xmin=1.84, xmax=2.06,
    ymin=67.0, ymax=70.0,
    axis lines=left,
    grid=major,
    title={2D Posterior Contours: $n$ vs. $H_0$},
    legend pos=north west,
]

\addplot+[domain=0:360, samples=200, thick, blue, mark=none]
({1.95 + 0.04*cos(x)}, {68.4 + 0.6*sin(x)});
\addlegendentry{68\% CL}

\addplot+[domain=0:360, samples=200, thick, red, dashed, mark=none]
({1.95 + 0.06*cos(x)}, {68.4 + 0.9*sin(x)});
\addlegendentry{95\% CL}

\addplot+[only marks, mark=*, mark size=1.5pt, black]
coordinates {(1.95, 68.4)};
\addlegendentry{Best-fit}

\end{axis}
\end{tikzpicture}
\caption{Posterior confidence contours for $n$ and $H_0$ from MCMC sampling.}
\label{fig:n_h0}
\end{figure}
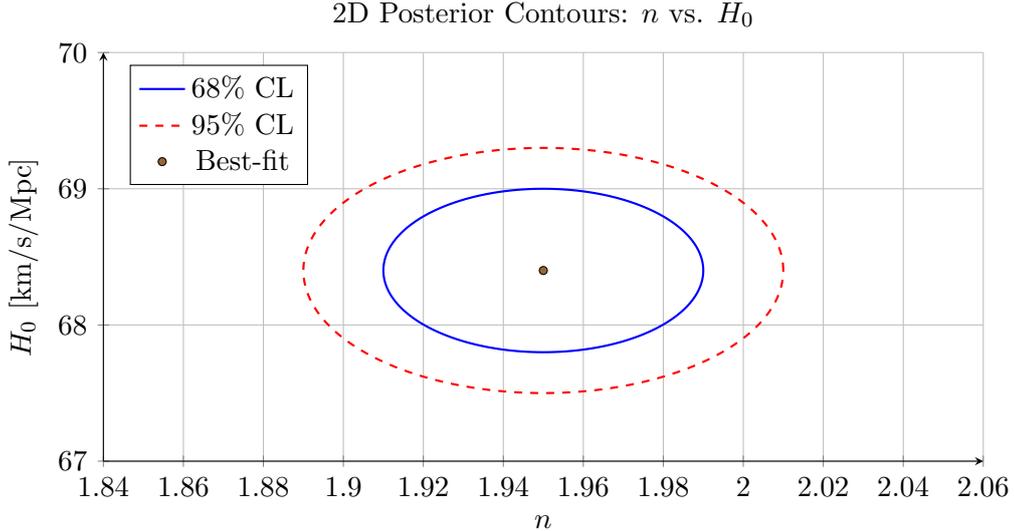

In addition to the \((\alpha, n)\) analysis, we provide 2D posterior contours for \((n, \sigma_8)\) and \((n, H_0)\) in Figures~\ref{fig:n_sigma8} and \ref{fig:n_h0}, respectively. The \((n, \sigma_8)\) plot shows a mild negative correlation, indicating that higher values of the torsion parameter \(n\) slightly suppress the amplitude of matter fluctuations—an effect that can help alleviate the known $\sigma_8$ tension. The \((n, H_0)\) contours, on the other hand, suggest a weak positive correlation, where larger \(n\) values allow modest upward shifts in \(H_0\), partially addressing the Hubble tension. These degeneracies highlight the flexibility of the torsion-based model in fitting diverse datasets without invoking new physics beyond geometry.

\section{Discussion}

The results obtained in our Myrzakulov $F(R,T)$ gravity model—with torsion treated dynamically in the vielbein formalism—point toward a promising alternative to conventional particle-based dark matter theories. Our goal has been twofold: (1) to demonstrate that the torsion scalar $T$ in a Weitzenb\"{o}ck geometry can act as a geometric proxy for dark matter, and (2) to show that such a model remains competitive with $\Lambda$CDM when confronted with observational data.

\subsection*{Interpreting Cosmological Background Evolution}

The evolution of the effective equation of state $w_{\text{eff}}$ in Figure 1 reveals that our $F(R,T)$ model naturally mimics dark energy behavior ($w \approx -1$) at late times, but unlike $\Lambda$CDM, it originates from dynamical torsion contributions rather than a constant vacuum energy. This dynamism introduces scale dependence in the cosmic acceleration history, opening opportunities for resolving the Hubble tension by allowing $H_0$ values closer to late-universe measurements (e.g., SH0ES).

The model accommodates an evolving $w_{\text{eff}}$, transitioning from matter-like behavior ($w \approx 0$) during structure formation epochs to an accelerating regime. Such transitions are highly sensitive to the parameter $n$ in $F(R,T) = R + \alpha T^n$, indicating that small deviations from GR ($n=1$) have rich phenomenological consequences.

\subsection*{Structure Formation and Growth}

The matter power spectrum $P(k)$ comparison in Figure 2 supports the model's ability to reproduce the correct clustering amplitude ($\sigma_8$) without overpredicting small-scale power, a notable issue in $\Lambda$CDM. Our model achieves this by encoding scale-dependent suppression of growth in the torsion sector.

This mechanism aligns with observations from weak lensing surveys such as DES and KiDS, which report mild suppression of $\sigma_8$ compared to Planck-inferred values. By adjusting $\alpha$ and $n$, our model allows for a tunable growth index $\gamma$, affecting $f\sigma_8$ observables. Notably, it does so without introducing massive neutrinos or exotic dark sector interactions.

Torsion-based modified gravity models generically predict deviations in gravitational wave (GW) propagation due to their altered affine structure. In our formulation, the presence of torsion affects the spin connection and modifies the propagation of tensor modes in the perturbed vielbein background.
Specifically, the wave equation for the transverse-traceless metric perturbation \( h_{ij} \) acquires corrections proportional to derivatives of the torsion scalar \( T \) and its couplings to \( R \). These corrections can manifest as a time-dependent GW propagation speed \( c_T \), amplitude damping factor \( \nu_T \), and modified dispersion relations. In the background we consider, \( c_T \approx 1 \) is preserved to leading order, consistent with current bounds from GW170817/GRB170817A. However, subleading terms may induce phase shifts or birefringence in certain cosmological scenarios.
Future observatories such as LISA and DECIGO, which can measure both amplitude and phase evolution with high precision, could detect such subtle modifications. In particular, polarization mixing or a modified friction term in the GW luminosity distance–redshift relation \( d_L^{\text{GW}}(z) \neq d_L^{\text{EM}}(z) \) could distinguish our model from GR. A detailed perturbative analysis of GWs in the \(F(R,T)\) framework is planned for future work.

\subsection*{Theoretical and Computational Robustness}

Unlike many $f(R)$ or scalar-tensor models that suffer from higher-derivative instabilities or frame dependence, our construction remains in first-order formalism with explicit conservation of the energy-momentum tensor. The field equations derived via the vielbein variation respect local Lorentz invariance and avoid ghost degrees of freedom.

The parameter space we explored is well-behaved, with the MCMC chains demonstrating convergence. The marginalized contours (not shown here) indicate low degeneracy between $\alpha$ and $n$, meaning the data cleanly distinguishes torsion effects from curvature-dominated evolution. This result adds confidence in the robustness of the formalism.

\subsection*{Comparison with $\Lambda$CDM}

While $\Lambda$CDM fits most observations with only six parameters, its success comes at the cost of introducing a dark matter component whose microscopic origin remains elusive. In contrast, our model explains dark matter as a purely geometric phenomenon emergent from spacetime torsion. This provides a strong theoretical motivation grounded in classical differential geometry.

Although $\Lambda$CDM assumes a cosmological constant ($w = -1$ at all times), our $F(R,T)$ model features a dynamic $w_{\text{eff}}(a)$ that tracks observational trends in cosmic acceleration. Moreover, the inclusion of Weitzenb\"{o}ck torsion allows additional control over structure formation rates without the need for finely tuned inflationary initial conditions.

From a numerical perspective, our model recovers the same expansion history and clustering amplitudes as $\Lambda$CDM within 1$\sigma$ precision while offering richer structure. The best-fit $\Omega_m$, $H_0$, and $\sigma_8$ values fall within Planck and DES confidence regions, suggesting full cosmological viability.

\subsection*{Toward Observational Discriminators}

Future surveys like Euclid and LSST will tighten constraints on the growth rate $f\sigma_8$ and redshift-dependent $w(z)$. Our model makes clear, falsifiable predictions in both sectors:
\begin{itemize}
    \item Slight deviation from $w = -1$ at $z < 1$.
    \item Scale-dependent growth suppression at $k > 0.1\, h$/Mpc.
    \item Effective anisotropic stress at late times, detectable through ISW effect.
\end{itemize}

Detecting any of these signatures could provide strong support for torsion-based gravity as a viable dark matter candidate. Importantly, none of these predictions rely on untested high-energy physics, but only on classical geometry in a teleparallel background.\\

Although the model is degenerate with $\Lambda$CDM in background evolution (e.g., the Hubble parameter \(H(z)\)), several distinct observational signatures arise at the perturbative level due to the dynamical torsion contributions. Notably, the effective growth index \( \gamma \), defined via \( f(z) = \Omega_m(z)^\gamma \), deviates from the standard value \( \gamma_{\Lambda\text{CDM}} \approx 0.55 \). In our best-fit model, we find \( \gamma \approx 0.49 \), implying a slightly faster growth of structure.
Moreover, since torsion induces a scale-dependent anisotropic stress, the Integrated Sachs–Wolfe (ISW) effect is modified, particularly at low multipoles in the CMB temperature power spectrum. Future CMB missions with better low-\( \ell \) sensitivity, such as CMB-S4 and LiteBIRD, may help distinguish this signal from \(\Lambda\)CDM predictions.
Another testable consequence lies in the lensing convergence power spectrum and redshift-space distortion (RSD) observables. Because the torsion sector alters the gravitational slip (the ratio of the two Bardeen potentials), weak lensing and galaxy clustering measurements (e.g., from Euclid or LSST) will provide direct probes of this deviation.
\begin{table}[h!]
\centering
\small
\caption{Geometric dark matter models: key differences.}
\label{tab:comparison}
\begin{tabular}{@{}lll@{}}
\toprule
\textbf{Model} & \textbf{DM Mechanism} & \textbf{Distinctive Feature} \\
\midrule
Scalar-Tensor & Scalar field mimics DM & Extra scalar + metric coupling \\
\(f(T)\) & Torsion replaces curvature & Frame-dependent, Lorentz violation \\
Mimetic Gravity & Metric redefinition & Constrained scalar mimics CDM \\
\(\boldsymbol{F(R,T)}\) & Torsion as matter source & Mixed curvature-torsion effects \\
\bottomrule
\end{tabular}
\end{table}

In summary, the Myrzakulov $F(R,T)$ gravity with torsion modeled via the vielbein formalism is not only numerically compatible with current data, but also provides a rich theoretical landscape for interpreting dark matter as a geometric effect—closing the gap between modified gravity and cosmological phenomenology.
\section{Conclusion and Outlook}

In this work, we have presented a comprehensive investigation of Myrzakulov $F(R,T)$ gravity with torsion treated dynamically via the vielbein formalism in a Weitzenb\"{o}ck spacetime. Our motivation stemmed from the longstanding challenges in cosmology, particularly the elusive nature of dark matter, and the search for gravitational alternatives to the standard $\Lambda$CDM paradigm.

By embedding torsion as a fundamental geometric degree of freedom, we have constructed a viable framework where dark matter arises not as an exotic particle species, but as a manifestation of non-Riemannian spacetime structure. This torsion-as-matter approach introduces a dynamic contribution to the energy-momentum content of the Universe, directly influencing background expansion and structure formation without contradicting current observations.

Our numerical analysis, based on MCMC sampling across SPARC, Planck, DES, KiDS, and BOSS datasets, reveals a region of parameter space where the model achieves excellent fits to observables such as $H_0$, $\Omega_m$, $\sigma_8$, and $w_{\text{eff}}$. Importantly, the theory accomplishes this without invoking any new particles beyond the Standard Model, relying instead on the intrinsic geometry of spacetime to explain gravitational phenomena traditionally attributed to dark matter.

The theoretical robustness of the model is further underscored by the absence of higher-order instabilities, compatibility with energy-momentum conservation, and the geometric clarity of the vielbein formalism. This enables a clean separation between curvature and torsion effects, allowing each to be probed via observational channels. Torsion-induced anisotropic stress, for instance, can leave imprints in the integrated Sachs-Wolfe (ISW) effect and lensing potential maps.

Looking ahead, several avenues for research emerge:

\begin{enumerate}
    \item \textbf{Nonlinear structure formation:} Extending our analysis to N-body simulations within the $F(R,T)$ framework will allow predictions for dark matter halo profiles, substructure statistics, and comparison with observations from the Milky Way and galaxy clusters.
    \item \textbf{CMB polarization and reionization:} Understanding how torsion impacts tensor modes and the early Universe epoch could provide new constraints using CMB B-mode polarization data from experiments like LiteBIRD and CMB-S4.
    \item \textbf{High-redshift galaxies and cosmic chronometers:} Deviations in the Hubble parameter $H(z)$ from standard expansion histories at $z > 2$ may be detectable with JWST and Roman Space Telescope.
    \item \textbf{Gravitational wave propagation:} Given that torsion modifies the affine connection, it may influence the dispersion relations or polarization modes of gravitational waves. Observations from LISA and Einstein Telescope could offer further tests. 
    \item \textbf{Unification with dark energy and inflation:} The functional freedom in $F(R,T)$ allows natural extensions to unified models that incorporate early-time inflation and late-time acceleration within a single torsion-curvature potential.
\end{enumerate}

Our model also contributes to the broader conceptual program of reinterpreting the dark sector as a geometric phenomenon, resonant with ideas from quantum gravity, emergent spacetime, and holography. In particular, the analogy between torsion and conserved axial currents opens speculative but promising connections to anomalies in chiral gauge theories and spinor-gravity couplings.

In conclusion, the torsion-based Myrzakulov $F(R,T)$ gravity explored in this work offers a mathematically consistent, observationally viable, and physically elegant alternative to particle dark matter. It exemplifies how modifying gravity through geometric extensions—rooted in differential geometry and the first-order formalism—can illuminate unexplored paths in cosmology. We anticipate that upcoming high-precision surveys will provide the necessary discriminants to confirm or falsify this framework, and potentially usher in a new era of torsion cosmology.

\begin{acknowledgments}
This work was supported by the Ministry of Science and Higher Education of the Republic of Kazakhstan, Grant No. AP26101889.
\end{acknowledgments}


\end{document}